\documentclass[11pt]{article}

\usepackage[dvips]{epsfig}
\usepackage{epsfig}
\usepackage{graphicx}
\usepackage{color}
\usepackage{fullpage, times}

\usepackage{latexsym}
\usepackage{amsmath}
\usepackage{amssymb}

\long\def\commabs #1\commabsend{}

\newcommand{\reals}{\hbox{$\rlap{\rm I} \> \kern-.2mm{\rm R}$}}

\newcommand{\hd}{\hat{\delta}}
\newcommand{\etal}{{\em et al.\ }}

\commabs

\commabsend

\newcommand{\dist}{\mbox{\bf dist}}

\newcommand{\eps}{\epsilon}

\def\inline#1:{\par\vskip 7pt\noindent{\bf #1:}\hskip 10pt}
\def\Proof{\par\noindent{\bf Proof:~}}
\def\Proofof#1{\par\noindent{\bf Proof of #1:~}}

\def\tO{\tilde{O}}

\newtheorem{theorem}{Theorem}[section]
\newtheorem{lemma}[theorem]{Lemma}

\newtheorem{claim}[theorem]{Claim}

\newtheorem{definition}{Definition}

\def\inline#1:{\par\vskip 7pt\noindent{\bf #1:}\hskip 10pt}
\def\Proof{\par\noindent{\bf Proof:~}}
\def\blackslug{\hbox{\hskip 1pt \vrule width 4pt height 8pt
    depth 1.5pt \hskip 1pt}}

\def\QED{\quad\blackslug\lower 8.5pt\null\par}

\renewcommand{\paragraph}[1]{\par\noindent\textbf{#1}}

\newcommand{\Heap}{\mbox{\bf Heap}}

\def\hd{\mbox{\tt heavy\_dist}}
\def\ld{\mbox{\tt light\_dist}}

\begin{document}

\title{Dynamic Decremental Approximate Distance Oracles with $(1+\epsilon, 2)$ stretch}

\author{Ittai Abraham
\thanks{Microsoft Research Silicon Valley, Mountain View CA, USA.
Email: {\tt \{ittaia,schechik\}@microsoft.com}.}
\and
Shiri Chechik~$^*$
}

\date{\today}
\maketitle

\begin{abstract}
We provide a decremental approximate Distance Oracle that obtains stretch of $1+\epsilon$ multiplicative and 2 additive and has $\hat{O}(n^{5/2})$ total cost (where $\hat{O}$ notation suppresses polylogarithmic and $n^{O(1)/\sqrt{\log{n}}}$ factors).
The best previous results with $\hat{O}(n^{5/2})$ total cost obtained stretch $3+\epsilon$.
\end{abstract}

\thispagestyle{empty}
\maketitle
\newpage
\setcounter{page}{1}

\section{Introduction}

Dynamic graph algorithms are designed to maintain some functionalities on the network in the settings where the network changes over time.
This paper considers the problem of maintaining (approximate) shortest paths in the dynamic setting, where edges are being deleted and added to the graph.

\paragraph{Dynamic distance oracles:}

A \emph{dynamic distance oracle} (DDO) is a data structure that is capable of efficiently processing an adversarial sequence of delete, insert and distance query operations.
A \emph{delete} operation deletes a single edge from the graph.
An \emph{insert} operation adds a single edge to the graph.
A \emph{query} operation receives a pair of nodes and returns a distance estimation.
We say that a dynamic algorithm is \emph{decremental} if it handles only deletion operations, \emph{incremental} if it handles only insertion operations, and \emph{fully dynamic} if it handles both.
A dynamic \emph{approximate distance oracle} has \emph{stretch} $k$ if the returned distance estimation for every pair of nodes is at least the actual distance between them and at most $k$ times their actual distance.
A  \emph{single-source} dynamic distance oracle (SSDDO) has a fixed source $s$ and all distance queries must involve the source $s$.
One can obtain a dynamic distance oracle by simply constructing a dynamic single-source distance oracle for every possible source.

Even for single-source decremental dynamic distance oracles we do not know of any non-trivial bounds on worst-case operation costs. So it is natural to consider amortized costs as the next best measure. The \emph{amortized cost} of a dynamic distance oracle is the average cost of a sequence of $m$ operations taken over all possible adversarial sequences and all possible graphs with $n$ vertices and $m$ edges. Note that simply running Dijkstra's algorithm on queries (and trivially updating the graph data structure on delete and insert operations) gives a $\tilde{O}(m)$ amortized cost DDO for exact distances.
The \emph{worst case query time} is the bound on the cost of any query. This bound is important when one expects significantly more query operations relative to delete and insert operation.

The dynamic distance oracle problem (with its various variations) has received a lot of attention in the last three decades.  We survey some of the main results:

\paragraph{Exact Single-Source DDOs:}
Even and Shiloach, in 1981, presented a decremental SSDDO for undirected, unweighted graphs with $O(m)$ amortized cost and $O(1)$ query time with stretch 1 (exact distances). A similar scheme was independently found by Dinitz \cite{Di06}.
Later, King \cite{King99} generalized this result to directed graphs.
The naive implementation of the dynamic distance oracle of \cite{King99} requires in the worst case $O(n^3)$ memory.
King and Thorup \cite{KiTh01} showed a technique that allows implementing a dynamic distance oracle using the algorithm of \cite{King99} with only $O(n^{2.5})$ ($O(n^{2} \sqrt{nb})$ memory, where $b$ is the maximal edge weight).

Roditty and Zwick \cite{RoZw04b} showed that incremental and decremental SSDDO for unweighted graphs are at least as hard as several basic problems such as Boolean matrix multiplication and the problem of finding all edges in a given graph that are part of a triangle.

\paragraph{Exact DDOs:}
The problem of exact DDO was extensively studied.
Ausiello \etal \cite{AuItMaNa91} presented an incremental DDO for weighted directed graphs with amortized cost $O(n^3 \log{n}/m)$ and $O(1)$ query time.
Henzinger and King showed a decremental DDO for weighted directed graphs with amortized cost $\tilde{O}(n^2/t + n)$ and $O(t)$ query time.

Later, King \cite{King99} presented a fully dynamic DDO for unweighted graphs with  amortized cost $\tilde{O(n^{2.5})}$ and $O(1)$ query time.
Demetrescu and Italiano \cite{DeIt06} presented a fully dynamic DDO for directed weighted graph with amortized cost $\tilde{O}(n^{2.5} \sqrt{S})$, where $S$ is the possible number of different weight values in the graph.

Demetrescu and Italiano~\cite{DeIt04}, in a major breakthrough devised a fully dynamic exact DDO for directed
general graphs with non negative edge weights, with amortized cost $\tilde{O}(n^2)$.
Thorup~\cite{Thorup04-Fully} later extended the
result to negative edge weights and slightly improved the update time.
Thorup~\cite{Thorup05} also considered the worst case update time and presented fully dynamic DDO with
worst case update time $\tO(n^{2.75})$.
Baswana et al.~\cite{BaHaSe02} devised a decremental DDO  for unweighted directed graphs and
amortized cost $\tO(n^3/m)$.

\paragraph{Approximate DDOs, incremental-only and decremental-only:}
The dynamic distance oracle problem was also studied when approximated distances are allowed. We begin with the incremental-only and decramental-only results. Baswana \etal \cite{BaHaSe02} presented a decremental DDO for unweighted graphs with amortized cost $\tilde{O}(n^{2} /\sqrt{m})$, $O(1)$ query time and $(1+\eps)$ stretch.
Later, Baswana \etal \cite{BaHaSe03} presented several decremental algorithms for undirected graphs.
They presented stretch 3 decremental DDO with amortized cost $\tilde{O}(n^{10/9})$,
stretch 5 decremental DDO with amortized cost $\tilde{O}(n^{14/13})$,
and stretch 7 decremental DDO with amortized cost $\tilde{O}(n^{28/27})$.
Roditty and Zwick~\cite{RoZw04,RoZw12} presented extremely efficient distance oracles for the only
incremental and for the only decremental cases. Each has amortized cost $\tO(n)$,
$(1+\eps)$ stretch and $O(1)$ query time. 
In a recent breakthrough Bernstein \cite{Be13} obtained similar bounds for directed weighted graphs.
Roditty and Zwick~\cite{RoZw04,RoZw12} also presented a second decremental algorithm with amortized cost $\tO(n)$,
$(2k-1)$ stretch and $(k)$ query time that uses a space of $O(m+n^{1+1/k})$ (rather than a space of $O(mn)$).
Bernstein and Roditty~\cite{BeRo11} later presented a decremental DDO for unweighted undirected graphs with
$(2k-1+\eps)$ stretch, $O(k)$ query time and amortized cost $\tO(n^{2+1/k+O(1)/\sqrt{\log{n}}} /m)$.
In the same paper Roditty and Bernstein also presented a very efficient decremental SSSP for unweighted undirected graphs with amortized cost $\tilde{O}(n^{2+O(1/\sqrt{\log{n}})}/m)$, $(1+\eps)$ stretch and constant query time.

\paragraph{Fully dynamic approximate DDOs:}
For the fully dynamic approximate DDO problem the following results were achieved.
King \cite{King99} presented a fully dynamic DDO with amortized cost $\tilde{O}(n^{2})$, $O(1)$ query time and $(1+\eps)$ stretch.
Roditty and Zwick~\cite{RoZw04,RoZw12} presented a fully dynamic DDO for any fixed $\epsilon, \delta >0$ and every $t \leq m^{1/2-\delta}$,
with expected amortized cost of $\tilde{O}(mn/t)$ and worst case query time of $O(t)$ and $(1+\eps)$ stretch.
Note that as $t \leq m^{1/2-\delta}$, the best amortized cost that can be achieved using this algorithm is $\Omega(m^{1/2+\delta}n) > \Omega(m)$.

Later, Bernstein~\cite{Be09} presented fully dynamic DDO with $O(\log\log\log{n})$ query time, $2+\eps$ stretch and  $\tilde{O}(m n^{O(1)/\log{n}})$ amortized cost.

\subsection{Our contributions}


We construct a decremental approximate DDO that obtains stretch of $1+\epsilon$ multiplicative and 2 additive. Note that this is at most $2+\epsilon$ multiplicative since we can answer exactly on edges. Our decremental approximate DDO has only $\hat{O}(n^{5/2})$ total cost\footnote{$\hat{O}(f(n)) = f(n) n^{O(1)/\sqrt{n}}$ be a crude way to suppress poly-log and $n^{O(1)/\sqrt{\log{n}}}$ factors.}. Previously the best results for decremental approximate DDO with $\hat{O}(n^{5/2})$ total cost obtained stretch $3+\epsilon$ \cite{{BeRo11}}. 

\begin{theorem}\label{thm:main}
One can maintain a decremental dynamic distance oracle, of size $O(n^{5/2})$ with $(1+\epsilon,2)$ stretch, constant query time, and total cost  of $\hat{O}(n^{5/2})$.
\end{theorem}

\paragraph{Additional related work:}
A related notion of dynamic distance oracle is that of distance oracles supporting a fixed number of failures.
A distance oracle supporting a single edge failure with exact distances, $\tilde{O}(n^2)$ size and  $O(\log
n)$ query time was presented in~\cite{DT02}.
This was later generalize to handle a single edge or vertex failures \cite{DT02} and then to dual failures ~\cite{DP09}.
Approximate dynamic distance oracles supporting multiple edge failures was presented in~\cite{CLPR10}.

A more relaxed version of the dynamic distance oracle is that of the dynamic connectivity oracle.
In this problem it is required to answer connectivity queries rather than distance queries.
It is not hard to see that any result on dynamic distance oracle with any stretch automatically implies dynamic connectivity oracle with the same bounds.
The problem of dynamic connectivity oracle was extensively studied.
Dynamic connectivity oracle with poly-log amortized update time were first introduced by Henzinger and King \cite{HeKi99}
(see \cite{HeTh97,HoLiTh01,Thorup00,PaDe04} for further improvements and lower bounds).

The problem of constructing dynamic connectivity problem with worst case update time was also considered.
Frederickson \cite{Fr85} introduced dynamic connectivity oracle with $O(\sqrt{m})$ update time.
The sparsification technique of Eppstein et. al. \cite{EppGaItNi92,EppGaItNi97} improved the update time to $O(\sqrt{n})$.

P\v{a}tra\c{s}cu and Thorup~\cite{PaTh07} considered the connectivity problem in a restricted model where all edge deletions occur in one bunch and after
the deletions, distance queries arrived.
They presented a data structure of size $O(m)$ such that given a set $F$ of
of $f$ edge failures and two nodes $s$ and $t$, can decide if $s$ and $t$ remain connected in time $\tO(f)$.

Duan and Pettie \cite{DuPe10} later considered the same problem for vertex failures and presented
a data structure of size $\tilde{O}(f^{1-2/c} m  n^{1/c - 1/(c \log{2f})})$, $\tilde{O}(f^{2c+4})$ update time,
and $O(f)$ query time, where $c$ is some integer and $f$ is the number of vertex failures occurred.

In a recent breakthrough, Kapron \etal \cite{KaKiMo13} showed a construction for fully dynamic distance oracle with poly-log worst case update and
query time.
%
%

\section{Preliminaries}

\subsection{Existing Decremental SSSP algorithms}

Our algorithm uses the decremental SSSP algorithm of King \cite{King99} as an ingredient and modify it.
The properties of King's algortihm are summarized in the following theorem.

\begin{theorem}\cite{King99}
\label{thm:King}
Given a directed graph with positive integer edge weights, a source node $s$ and a distance $d$,
one can decrementally maintains a shortest path tree $T$ from $s$ up to distance $d$ in total time $O(md)$.
Moreover, given a node $v$, one can extract in $O(1)$ time $\dist(v,s)$ in case $v \in T$ or determine that $v \notin T$.
\end{theorem}

King's algorithm starts by constructing a shortest path tree $T$ rooted at $s$.
Each time an edge $(x,y)$ is deleted, where $x$ is in the same connected component as $s$ in $T \setminus e$, an attempt is made to find a substitute edge to $y$ that does not increase
the distance from $s$ to $y$. If such edge is found then the recovery phase is over.
Note that in this case the distances from $s$ to $y$ and to all nodes in $y$'s subtree are unchanged.
In case no such edge found, the best edge is chosen, i.e., the edge that connect $y$ on the shortest path possible.
The process is continued recursively on all $y$'s children.
The crucial property of this algorithm is that it explores the edges of a node $v$ only when the distance from $s$ to $v$ increases.
This gives a total running time of $O(md)$ as the distance from $s$ to a node $v$ may increase at most $d$ times before exceeding $d$.

Our algorithm also uses as an ingredient the efficient construction of Bernstein and Roditty~\cite{BeRo11} for maintaining a $(1+\eps)$ decremental SSSP.
The input of the algorithm is an undirected unweighted graph and a source node $s$.
The algorithm decrementally maintains a $(1+\eps)$ shortest path tree $T$ from $s$ in total time $\hat{O}(n^{2})$.
More specifically, Roditty and Bernstein showed the following.
They showed how to maintain a $(1+\eps/2, n^{\frac{\sqrt{6/\eps}}{\sqrt{\log{n}}}})$ emulator $H$ in time $\hat{O}(m)$.
Let $\zeta = n^{\frac{\sqrt{6/\eps}}{\sqrt{\log{n}}}}$ and $\beta = (2/\eps)\zeta$.
They show that if $\dist(x,y) \geq \beta$ then $\dist(x,y,H) \leq (1+\eps)\dist(x,y)$.
In addition, they show how to maintain a tree $T(s)$, where the distances $\dist(s,x,T(s)) = \dist(s,x,H)$ for every $x \in V$.
In order to get rid of the additive term for short distances they handle short distances separately.
Let $\dist^{BR}(s,x)$ be the estimated distance returned by Roditty and Bernstein's decremental SSSP algorithm.
Let $H$ be the emulator in the construction of Bernstein and Roditty \cite{BeRo11}.
We summarize the properties we need from Bernstein and Roditty's construction in the following theorem.

\begin{theorem}\cite{BeRo11}
\label{thm:RB}
For a given graph $G$ and a node $s$, one can maintain a decremental $(1+\eps)$ emulator $H$ and
a shortest path tree $T = T(s)$ from $s$ in $\hat{O}(n^{2})$ total time with the following properties: \\
(1) The graph $H$ is a $(1+\eps/2, n^{\frac{\sqrt{6/\eps}}{\sqrt{\log{n}}}})$ emulator, namely, for every two nodes $x$ and $y$, $\dist(x,y) \leq \dist(x,y,H) \leq (1+\eps/2)\dist(x,y) + n^{\frac{\sqrt{6/\eps}}{\sqrt{\log{n}}}})$.  \\
(2) If $\dist(x,y) \geq \beta$ then $\dist(x,y,H) \leq (1+\eps)\dist(x,y)$. (this follows directly from $(1)$ by straightforward calculations). \\
(3) For every $x \in V$: $\dist(s,x,T(s)) = \dist(s,x,H)$
\end{theorem}

For our construction we also need the following additional property from the emulator.

\begin{lemma}
\label{lem:emulator-very-long-dis}
Consider two nodes $x$, $y$ and $z$, if $\dist(x,y) \geq 8 \beta/\eps$ and $z$ is at distance at most $\beta$ from some node on $P(x,y)$ then
$\dist(x,z,H) + \dist(y,z,H) \leq (1+\eps)\dist(x,y)$.
\end{lemma}


\section{Decremental with $\hat{O}(n^{5/2})$ total update time}
\label{sec:2-add-dec}

In this section we present a new decremental all-pairs shortest paths algorithm with $\hat{O}(n^{5/2})$ total update time, with
a multiplicative stretch of $1+\eps$ and additive stretch of 2.
In fact the stretch is the maximum between a multiplicative $1+\eps$  and additive stretch 2, namely, let $\hat{d}(x,y)$ be the reported distance, then
$\dist(x,y) \leq \hat{d}(x,y) \leq \dist(x,y) + \max\{\eps\dist(x,y),2\}$. For simplicity, we present a scheme that guarantees the following  $\hat{d}(x,y) \leq O(1+\eps)(\dist(x,y)+2)$ and with query time $O(\log\log{n})$. We later explain the slight modifications to improve the guarantee to $\hat{d}(x,y) \leq \dist(x,y) + \max\{\eps\dist(x,y),2\}$ and how to reduce the query time to constant.

We say that a node is {\em heavy} if it's degree is larger than $n^{1/2}$ or {\em light} otherwise.
Let $P(s,t)$ be a shortest path from $s$ to $t$.
Let $\hd(s,t)$ be the minimal distance between $s$ and $t$ that goes through some heavy node, namely, $\hd(s,t) = \min\{\dist(s,x) + \dist(x,t) \mid x~is~heavy\}$.
Let $\ld(s,t)$ be the length of the shortest path between $s$ and $t$, where all nodes on that path are light.
Let $\dist_v(x,y)$ be the length of the shortest path from $x$ to $y$ that goes through $v$, namely, $\dist_v(x,y) = \dist(x,v) + \dist(v,y)$.
Let $\dist_Q(x,y)$ be the minimal distance $\dist_v(x,y)$ for some $v \in Q$.
Let $\dist_v^{BR}(x,y)$ be the distance $\dist^{BR}(v,x) + \dist^{BR}(v,y)$. 

Previous decremental algorithms used dynamic SSSP as an ingredient by including all nodes in the tree through the entire execution of the algorithm (or all nodes up to some distance).
We maintain decremental SSSP that includes only some of the nodes, and nodes may be added to the tree at some later stage of the algorithm.
In fact, some nodes may be added and removed from the tree many times during the algorithm.
Roughly speaking, we would like to add to the tree $T(v)$ only nodes whose shortest path to $v$ does not contain any heavy nodes.
This raises several difficulties. Note that just ignoring heavy nodes is not enough.
There may be a shortest path from $x$ to $v$ that contains a heavy node, but also a different longer path from $x$ to $v$ that does not go through any heavy node.
If we are not careful, we may add the node $x$ to the tree $T(v)$ on a path that is not the shortest.
As the graph changes at some point there might be no more heavy nodes on $P(x,v)$ anymore. At this point we may want that the distance $\dist(x,v,T(v))$ will be optimal or close to optimal.
This may result in shortening the distance from $x$ to $v$ in $T(v)$, which may be problematic as usually decremental SSSP algorithms rely on the fact that distances can only increase and thus it is possible it bound the number of times the distances change.
Therefore we need to be careful and add $x$ to $T(v)$ only if the shortest path $P(x,v)$ does not contain any heavy nodes.
Moreover, note that as $P(x,v)$ changes over time, it might changes between having heavy nodes to not having many times.
So the algorithm may  need to add and remove $v$ from the tree many times.

Loosely speaking, the algorithm maintains heavy distances by sampling a set $Q$ of $\tilde{O}(n^{1/2})$ nodes and maintaining $(1+\eps)$ shortest paths distances from all nodes in $Q$.
This is done using the construction of Roditty and Bernstein~\cite{BeRo11}.
In order to estimate $\dist_Q(x,y)$ the algorithm stores the distances $\dist_q^{BR}(x,y)$ for $q\in Q$ in a heap and updates the heap each time $\dist_{BR}(x,q)$
or $\dist_{BR}(y,q)$ changes by a $(1+\eps)$ factor.
In order to handle light distances the algorithm picks sets $S_i$ of $\tilde{O}(n/2^i)$ nodes and maintain a shortest paths trees $T(s)$ from each node $s \in S_i$ up to distance $2^i$, where the goal is to include only nodes $x$ such that their shortest path $P(s,x)$ does not include heavy nodes.
In order for the algorithm to determine if the path $P(s,x)$ contains heavy nodes, the algorithm uses the approximated distances for $\dist_Q(s,x)$.
Some difficulties arise from the fact that we don't have the exact distances $\dist_Q(s,x)$ but rather approximated ones.
In order to be able to maintain the shortest path trees from every $s \in S_i$ with small update time, we need to make sure that we do not decrease distances.
The entire analysis of King's algorithm  \cite{King99} relies on the
crucial property that distances between every two nodes can be increased at most $d$ times before exceeding the distance $d$.
In our case since we only have approximated distances for $\dist_Q(s,x)$, we cannot be sure if a path $P(s,x)$ contains a heavy node or not.
We thus need to be more strict in the decision to add a node to $T(s)$.
We need to maintain the property that if $y \in P(s,x)$ was not added to $T(s)$ then $x$ will not be added to $T(s)$ as well.
In order to do that we exploit the fact that the distances $\dist_{BR}(x,q)$ represents distances from an emulator $H$.
Thus, if $y \in P(s,x)$ was not added to $T(s)$ since there is a good alternative path $P_1$ that goes through an heavy node then
since $H$ also contains a good alternative path $P_2$ from $y$ to $x$, we get that by concatenating these paths there is a
good alternative path from $x$ to $s$ that goes through $Q$.
However some additional problems arise from the fact that $H$ is not really a $1+\eps$ emulator but rather has an additive stretch.
The emulator $H$ has a $1+\eps$ multiplicative stretch only for distances larger than $\beta$.
Our solution to bypass this problem is to store exact distances from $x$ to small ball around it and then
check if there is a good alternative path that consists of a short exact path and then a path from $H$.

In addition, for nodes $x\in V$ and $s\in S_i$ for some $1\leq i \leq \log{n}$
as will explained later on it is not enough to update
the distances $\dist_q^{BR}(x,s)$ for $q\in Q$ in the heap each time $\dist_{BR}(x,q)$
or $\dist_{BR}(y,q)$ changes by a $(1+\eps)$ factor. We will rather have a more refined heaps for nodes $x\in V$ and $s\in S_i$ that will be updated each time
$\dist_q^{BR}(x,s)$ increases. In order to do this efficiently these refine heaps maintain only distances up to $2^i$.

Consider the tree $T$ rooted at some node $s$.
Let $v$ be a node such that $v \notin T$.
Let $d(v,s,B(T,1))$ be the minimal distance $\dist(s,x, T) +1$ such that $x$ is a neighbor of $v$ in $G$.

\paragraph{The algorithm:}


We now describe the different components in our data structure.

\textbf{The first component} is a subset $Q$ of the vertices obtained by sampling every node independently with probability $c \ln{n}/n^{1/2}$, for some constant $c$.


\begin{claim}
\label{claim:Q-size}
The expected size of the set $Q$ is $\tilde{O}(n^{1/2})$.
\end{claim}

\textbf{The second component} is a collection of subsets $S_i$ of the nodes for every $1 \leq i \leq \log{n}$, obtained as follows.
The set $S_i$ is obtained by sampling every node independently with probability $\min\{\frac{c \ln{n}}{\eps 2^i},1\}$.


\begin{claim}
For every $1 \leq i \leq \log{n}$,
the expected size of the set $S_i$ is $\min\{\frac{c \ln{n} n}{\eps 2^i},n\}$.
\end{claim}

Note that the number of considered graphs during the entire running of the algorithm is $m$ (as there are $m$ deletions from the graph).
The following lemma shows that with high probability for every considered graph some useful properties occur.

\begin{lemma}
\label{lem:heavy}
With probability $1 -3/n^{c-3}$, for every considered graph $G'$ during the entire running of the algorithm, the following happens: \\
(1) for every heavy node $v$, $\Gamma(v) \cap Q \neq \emptyset$, where $\Gamma(v)$ is the set of neighbours of $v$. \\
(2) for every vertex $v$ and every index $i$ such that $1\leq i\leq \log{n}$ and such that there exists a node $z$
such that $\dist(v,z,G) \geq \eps 2^i$: $S_i \cap B(v,\eps 2^i,G) \neq \emptyset$. \\
(3) for every vertex $v$ such that $|B(v,\beta)| \geq n^{1/2}$: $Q \cap B(v,\beta,G) \neq \emptyset$.
\end{lemma}

For the rest of the proof we assume that Lemma \ref{lem:heavy} holds for all versions of the graph.

\textbf{The third component}, hereafter referred to as $Exact_Q$, relies on component $Q$ and is as follows.
For every node in $q \in Q$, maintain an exact decremental shortest path tree up to distance $8\beta/\eps$ using King's algorithm  \cite{King99}.
Using $Exact_Q$ for every $v\in V$ and $q \in Q$, one can determine in constant time if $\dist(v,q) \leq 8\beta/\eps$ and if so extract $\dist(v,q)$.


\begin{claim}
Maintaining $Exact_Q$ takes $\hat{O}(n^{1/2} \cdot m) \leq \hat{O}(n^{5/2})$ total time.
\end{claim}
\Proof
By Claim \ref{claim:Q-size} the expected size of $Q$ is $\tilde{O}(n^{1/2})$.
For every node $q \in Q$ maintaining the shortest path tree up to distance $8\beta/\eps$ takes $O(\beta m)$ total time.
The claim follows.
\QED

\textbf{The forth component}, hereafter referred to as $BR_Q$, relies on component $Q$ and is as follows.
For every node $q \in Q$, maintain a $(1+\eps)$-approximate decremental SSSP using the algorithm of Roditty and Bernstein~\cite{BeRo11}.
Recall that the total update time for maintaining Roditty and Bernstein~\cite{BeRo11} data structure is $\hat{O}(n^2)$, we thus have the following.

\begin{claim}
Maintaining $BR_Q$ takes $\hat{O}(n^{5/2})$ total update time.
\end{claim}

\textbf{The fifth component}, hereafter referred to as ${\cal H}^{1}$ relies on components $Q$ and $Exact_Q$.
The goal of this component is to maintain $\dist_Q(x,y)$ exactly for short distances.

The component is done as follows.
For every nodes $x,y\in V$ do the following.
If $\dist_Q(x,y) \leq 8 \beta/\eps$ then the distance $\dist_Q(x,y)$ is maintained exactly.
This is done by maintaining a heap $\Heap^{(1)}_{(x,y)}$ containing all values $\Heap^{(1)}_{(x,y)}[q] = \dist_q(x,y)$ such that $\dist_q(x,y) <  8 \beta/\eps$.
The algorithm updates the heap each time $\dist(x,q)$ or $\dist(y,q)$ increases.
Let $\min(\Heap^{(1)}_{(x,y)})$ be the minimal value in $\Heap^{(1)}_{(x,y)}$ or infinity in case $\Heap^{(1)}_{(x,y)}$ is null.

\begin{claim}\label{claim:maintining-exact}
Maintaining ${\cal H}^{1}$ takes $\hat{O}(n^{5/2})$ total update time.
\end{claim}

\textbf{The sixth component}, hereafter referred to as ${\cal H}^{1+\eps}$, relies on components $Q$ and $BR_Q$.
The goal of this component is to allow approximating the distances $\dist_Q(x,y)$ for every $x,y \in V$.
The main idea is to keep all distance $\dist_q^{BR}(x,y)$ in a heap. Ideally, each time one of $\dist^{BR}(q,x)$
and $\dist^{BR}(q,y)$ changes, the heap should be updated. However, this may take too long as $\dist^{BR}(q,x)$
and $\dist^{BR}(q,y)$ may change many times and moreover these distances may also decrease.
Thus instead we update the heap each time one of $\dist^{BR}(q,x)$ or $\dist^{BR}(q,y)$ increases by a factor of $(1+\eps)$.
We then show that this is enough to get a good estimation on $\dist_Q(x,y)$.

The component is done as follows.
For every pair of nodes $x$ and $y$ keep all distances $\{\dist_q^{BR}(x,y) \mid q \in Q\}$ in a minimum heap $\Heap^{(1+\eps)}_{(x,y)}$, where the key is $q$ and the value is $\dist_q^{BR}(x,y)$.
Let $\Heap^{(1+\eps)}_{(x,y)}[q]$ be the value of the key $q$ in the heap $\Heap^{(1+\eps)}_{(x,y)}$. Let $\min(\Heap^{(1+\eps)}_{(x,y)})$ be the minimum value in the heap.

For every two nodes $x\in V$ and $q\in Q$ store a distance $d_{last}(x,q)$ initially is set to $\dist(x,q)$.
Each time the distance $\dist^{BR}(q,x)$ increases the algorithm checks if $\dist^{BR}(q,x) \geq d_{last}(x,q) (1+\eps)$, if so the algorithm
updates the values $\Heap^{(1+\eps)}_{(x,y)}[q]$ for every node $y$ and set $d_{last}(x,q) = \dist^{BR}(q,x)$.

The next lemma shows that for every nodes $x,y$, $\min((1+\eps) \min(\Heap^{(1+\eps)}_{(x,y)}), \min(\Heap^{(1)}_{(x,y)}))$ is a good approximation on the heavy distance from $x$ to $y$.

\begin{lemma}
\label{lem:stretch-heavy}
For every nodes $x,y \in V$, $\dist(x,y) \leq \tilde{d}(x,y) \leq (1+\eps)^2(\hd(x,y)+2)$, where
$\tilde{d}(x,y) =  \min((1+\eps) \min(\Heap^{(1+\eps)}_{(x,y)}), \min(\Heap^{(1)}_{(x,y)}))$.
\end{lemma}

\begin{claim}\label{claim:maintain-heavy}
Maintaining ${\cal H}^{1+\eps}$ takes $\hat{O}(n^{5/2})$ total update time.
\end{claim}

\textbf{The seventh component}, hereafter referred to as ${\cal H}^{*,1+\eps}$ relies on components $Q$ and $BR_Q$.
The goal of this component is similar to the goal of the previous component with some subtle changes.
Approximating the heavy distances is useful for two main uses. The first use is for the distance queries.
The second use is for deciding if a node $v$ should be added to some tree $T(s)$ for $s \in S_i$ for $1\leq i \leq \log{n}$.
For the latter use it is not enough to update the heap each time
$\dist^{BR}(s,q)$ is increased or when $\dist^{BR}(q,x)$ is increased by a $1+\eps$ factor.
We rather need that the heap to contain the correct values of $\dist_q^{BR}(s,x)$, as otherwise there could be a case where
the value of $\dist_Q^{BR}(s,y)$ is more updated than the value $\dist_Q^{BR}(s,x)$ for some $y \in P(s,x)$.
Thus the value in the heap $\min(\Heap^{(1+\eps)}_{(s,y)}) = \dist_Q^{BR}(s,y)$ but
$\min(\Heap^{(1+\eps)}_{(s,x)}) < \dist_Q^{BR}(s,x)$ and we might decide to add $x$ to $T(s)$ but not $y$.

The component is done as follows.

For every node $x\in V$, index $1\leq i \leq \log{n}$ and $y \in S_i$.
Keep all distances $\{\dist_q^{BR}(x,y) \mid q \in Q , \dist_q^{BR}(x,y) \leq (1+\eps)2^i\}$ in a minimum heap $\Heap^{(*,1+\eps)}_{(x,y)}$.
$\Heap^{(*,1+\eps)}_{(x,y)}$ is similar to $\Heap^{(1+\eps)}_{(x,y)}$ with the slight difference that we
update $\Heap^{(*,1+\eps)}_{(x,y)}[q]$ when either $\dist^{BR}(q,x)$ is increased
or when $\dist^{BR}(q,y)$ is increased, rather than waiting until it increases by a factor of $(1+\eps)$.
Notice that the distance $\dist_z^{BR}(x,y)$ may also decrease, in that case the algorithm does not update $\Heap^{(*,1+\eps)}_{(x,y)}$.
When the distance $\dist_z^{BR}(x,y)$ exceeds $(1+\eps)2^i$,  remove $z$ from the heap $\Heap^{(*,1+\eps)}_{(x,y)}$ permanently.

\begin{claim}\label{claim:maintain-heavy-2}
Maintaining ${\cal H}^{*,1+\eps}$ takes $\hat{O}(n^{5/2})$ total update time.
\end{claim}

\textbf{The eighth component}, hereafter referred to as $KING-S-L$ (stands for King for small distances for light balls) relies on $Q$, $Exact_Q$ and ${\cal H}^{1}$.

The goal of this component is to overcome the fact that the emulator $H$ has an additive stretch.
Recall that we would like to make sure that if a node $y\in P(s,x)$ is not added to $T(s)$ then also $x$ is not added to $T(s)$.
If $H$ was indeed a $(1+\eps)$ emulator then note that if $\dist(s,y,H) \leq (1+\eps)\dist(s,y)$ then also
$\dist(s,x,H) \leq (1+\eps)\dist(s,x)$. To see this note that $\dist(x,y,H) \leq (1+\eps)\dist(x,y)$, therefore
$\dist(s,x,H) \leq \dist(s,y,H) +\dist(y,x,H) \leq (1+\eps)\dist(s,x)$.

But $H$ is not a $(1+\eps)$ emulator and it could be that $x$ and $y$ are very close to one another (less than $\beta$) and thus $H$ does not contain a $(1+\eps)$-shortest path between them.
Therefore it could happen that $\dist(s,y,H) \leq (1+\eps)\dist(s,y)$ but  $\dist(s,y,H) > (1+\eps)\dist(s,y)$.
To overcome this issue, we do the following.

First if the distance $\dist(x,Q) \leq \beta$ then we can show that $\dist_Q(s,x)$ can be well estimated by ${\cal H}^{1}$  and ${\cal H}^{1+\eps}$ for every $s \in V$.
Otherwise, if the distance $\dist(x,Q) > \beta$ then we maintain exact distances from $x$ to all nodes at distance $\beta$ from it.
As $\dist(x,Q) > \beta$ the ball $B(x,\beta)$ contains only light nodes and thus maintaining $B(x,\beta)$ and their distances to $x$ can be done efficiently.

Then in order to decide if $x$ should be added to $T(s)$ we check all distances $\dist(x,w) + \dist_Q(w,s,H)$ for all $w \in B(x,\beta)$.
Note that now if $x$ and $y$ are close (at distance less than $\beta$) then $y \in B(x,\beta)$ and we have the exact distance between them and thus we don't need to rely on $H$ that does not return a good approximation for close nodes.

Formally, the component is done as follows.
For every node $x$, if $\dist(x,Q) > \beta$ then maintain decremental shortest path tree from $x$ up to depth $\beta$ using King's algorithm  \cite{King99}.
Let $B(x,\beta)$ be all nodes at distance at most $\beta$ from $x$.

Note that it could be that in the beginning of the algorithm $\dist(x,Q) \leq \beta$ but at some point $\dist(x,Q) > \beta$.
At the point that $\dist(x,Q) > \beta$, the algorithm constructs the decremental shortest path tree from $x$ up to depth $\beta$.

\begin{claim}\label{claim:maintain-king}
Maintaining $KING-S-L$ takes $\hat{O}(n^{2})$ total update time.
\end{claim}

\textbf{The ninth and main component}, hereafter referred to as $KING-L$ (stands for King for light distances) relies on all previous eighth components as is done as follows.

Consider a tree $T$ rooted at $s$. The following is a key definition:

\begin{definition}[is not light for $(s,T)$]
We say that $v$ {\em is not light for $(s,T)$} if one of the following holds:\\
(1) $d(v,s,B(T,1)) \leq 8\beta/\eps$ and $d(v,s,B(T,1)) \geq \dist_Q(v,s) -2$
(recall that $d(v,s,B(T,1))$ is the minimal distance $\dist(s,x, T) +1$ such that $x$ is a neighbor of $v$ in $G$.); or\\
(2) $d(v,s,B(T,1)) > 8\beta/\eps$, $\dist(v,Q) \geq \beta$ and $d(v,T) \geq \dist(v,w) + \Heap^{(*,1+\eps)}_{(w,s)}/(1+\eps)$ and $w\notin T$ for some $w \in B(s,\beta)$; or\\
(3) $d(v,s,B(T,1)) > 8\beta/\eps$ and $d(v,s,B(T,1)) \geq \min(\Heap^{(*,1+\eps)}_{(v,s)})/(1+\eps)$.
\end{definition}

For every node $s \in S_i$, maintain $T(s)$ decremnetally according to the decremental algorithm of King \cite{King99},
with the following change.
When an edge $e$ is removed from the tree $T(s)$ do the following.
Update the tree $T(s)$ according to King's algorithm with the following change.
Recall that by King's algorithm operates as follows. Each time an edge $(x,y)$ is deleted, where $x$ is in the same connected component as $s$ in $T \setminus e$, an attempt is made to find a substitute edge to $y$ that does not increase the distance from $s$ to $y$.
If such edge is found then the recovery phase is over.
In case no such edge found, the best edge is chosen, i.e., the edge that connect $y$ on the shortest path possible.
The process is continued recursively on all $y$'s children.

Instead we do the following.
First find the best edge $e$ that connect $y$ on the shortest path possible.
If the path of $y$ does not increase then the recovery phase is over.
Otherwise, check if $y$ is not light in $(s, T(s))$ and if $y$ is not light in $(s, T(s))$ then do not add $y$ to $T(s)$ and continue recursively on $y$'s children.
If it is not the case that $y$ is not light in $(s, T(s))$ then add $y$ to $T(s)$ using $e$ and continue recursively on $y$'s children.
In addition, each time the distance $\min(\Heap^{(*,1+\eps)}_{(s,y)})$ increases we check if $y$ is not light for $(s,T(s))$, if not then add $y$ to $T(s)$ with the best edge possible.

The next lemma is crucial to our analysis and its proof is quite subtle.
Ideally, we would like that $T(s)$ would contain all nodes $v$ such that their shortest path from $s$ to $v$ does not go through an heavy node.
However, since we don't have exact distances we might not add some of these nodes to $T(s)$, in case $Heap^{(1+\eps)}_{(v,s)}$ is already a close enough estimation on $\dist(v,s)$.
The next lemma shows that if $v$ is added to $T(s)$ then the distance from $v$ to $s$ in $T$ is a shortest path.

By the next lemma we get that if a node is added to $T(s)$ then  it's path in $T(s)$ is the shortest.
This property is important as otherwise we might need to decrease the distance from $u$ to $s$ in $T(s)$ in the future.

\begin{lemma}
\label{lem:short}
If a node $u$ belongs to $T(s)$ for some $s\in S_i$ then $\dist(u,s,T(s)) = \dist(u,s)$.
\end{lemma}

The next lemma shows that maintaining $KING-L$ takes $\hat{O}(n^{5/2})$ total update time.

\begin{claim}\label{lem:nine-ok}
Maintaining $KING-L$ takes $\hat{O}(n^{5/2})$ total update time.
\end{claim}

Finally, \textbf{the tenth component}, hereafter referred to as $Pivots$ is done as follows.
The algorithm maintains for every node $v$ and index $i$ a close node $p_i(v) \in S_i$.
This can be achieved by storing in a heap $\Heap_i(v)$ all distances $\dist(v,s,T(s))$ for every node $s$ such that $s\in S_i$ and $v\in T(s)$.

\begin{claim}
Maintaining $p_i(v)$ for every $v\in V$ and $1\leq i \leq \log{n}$ takes $\hat{O}(n^{5/2})$ total update time.
\end{claim}
\Proof
Finally, maintaining for every node $v$ and index $i$ the node $p_i(v) \in S_i$ can be done
by storing in a heap $\Heap_i(v)$ all distances $\dist(v,s,T(s))$ for every node $s$ such that $s\in S_i$ and $v\in T(s)$ for $s \in S_i$.
It is not hard to verify that this can also be done in $\tilde{O}(n^{2})$ total time.
\QED

\paragraph{The query algorithm:}
The query algorithm given pair of nodes $s$ and $t$ is done as follows.
Find the minimal index $i$ such that $t \in T(p_i(s))$. 
Return $\min\{\dist(s,p_i(s), T(p_i(s))) + \dist(t, p_i(s), T(p_i(s))),
\min(\Heap^{(1)}_{(s,t)}), (1+\eps) \min(\Heap^{(1+\eps)}_{(s,t)})\}$.

The query algorithm can be implemented in $O(\log\log{n})$ time by invoking a binary search on the indices $i$ for $1 \leq i \leq \log{n}$.


\begin{lemma}
\label{lem:not-in-light}
Consider nodes $u\in V$ and $s \in S_i$ for some $1\leq i \leq \log{n}$.
If $u \notin T(s)$ then
$$\min \{ \min( \Heap^{(*,1+\eps)}_{(u,s)}), \min( \Heap^{(1)}_{(u,s)})\} \leq (1+\eps)^2(\dist(u,s)+2).$$
\end{lemma}
\Proof
Consider nodes $u$ and $s \in S_i$ for some $1\leq i \leq \log{n}$ such that $u \notin T(s)$. Let $T=T(s)$.
Note that $u$ was not added to $T$ since either $u$ is not light for $(s,T)$
 or some other node in $P(u,s)$ is not light for $(s,T)$.
Let $y$ be the first node on $P(u,s)$ that is not light for $(s,T)$.
We need to consider the different cases why $y$ is not light for $(s,T)$.

First we claim that $d(y,s,B(T,1)) = \dist(s,y)$. To see this, let $y_0$ be the node before $y$ on the path $P(s,u)$.
Note that $u \in T$. By Lemma \ref{lem:short} we have $\dist(s,y_0,T) = \dist(s,y_0)$.
Note also that $\dist(s,y) \leq d(y,s,B(T,1)) \leq  \dist(s,y_0,T) + 1 = \dist(s,y)$.
We get that $d(y,s,B(T,1)) = \dist(s,y)$.

The first case is when $d(y,s,B(T,1)) \leq 8\beta/\eps$ and $d(y,s,B(T,1)) \geq \dist_Q(y,s) -2$.
In this case we have $\dist_Q(u,s) \leq \dist(u,y) + \dist_Q(y,s) \leq \dist(u,y) + \dist(y,s) +2 = \dist(u,s)+2$.
In this case we get $ \min(\Heap^{(1)}_{(u,s)}) \leq \dist_Q(u,s) \leq \dist(u,s) +2$, as required.

Consider the second case where  $d(y,s,B(T,1)) > 8\beta/\eps$, $\dist(y,Q) \geq \beta$ and $d(y,s,B(T,1)) \geq \dist(y,w) + \min(\Heap^{(*,1+\eps)}_{(w,s)})/(1+\eps)$ and $w\notin T$ for some $w \in B(s,\beta)$.

$\min(\Heap^{(*,1+\eps)}_{(u,s)}) \leq (1+\eps)\dist_Q(u,s)  \leq (1+\eps)(\dist(u,y) + \dist_Q(y,s)) \leq (1+\eps)(\dist(u,y) + \dist(y,w) + \min(\Heap^{(*,1+\eps)}_{(w,s)})  \leq (1+\eps)(\dist(u,y)+(1+\eps)\dist(y,s)) \leq
(1+\eps)^2\dist(u,s)$.

Consider the third case where  $d(y,s,B(T,1)) > 8\beta/\eps$ and $d(y,s,B(T,1)) \geq  \min(\Heap^{(*,1+\eps)}_{(y,s)})/(1+\eps)$.

In this case we have
$ \min(\Heap^{(*,1+\eps)}_{(u,s)}) \leq (1+\eps)\dist_Q(u,s)  \leq (1+\eps)(\dist(u,y) + \dist_Q(y,s)) \leq (1+\eps)(\dist(u,y) + \min(\Heap^{(*,1+\eps)}_{(y,s)}) ) \leq
1+\eps)(\dist(u,y) + (1+\eps)\dist(y,s))\leq (1+\eps)^2\dist(u,s)$.
\QED

The next lemma shows that the distance returned by the query algorithm is within the desired stretch.

\begin{lemma}
The distance $\hat{d}(s,t)$ returned by the query algorithm satisfies $\dist(s,t) \leq \hat{d}(s,t) \leq (1+\eps)^{O(1)}(\dist(s,t) +2)$.
\end{lemma}
\Proof
We first show that $\dist(s,t) \leq \hat{d}(s,t)$.
In order to show this, we show that $\dist(s,t) \leq \dist(s,p_i(s), T(p_i(s))) + \dist(t, p_i(s), T(p_i(s)))$ and
$\dist(s,t) \leq (1+\eps)\min(\Heap^{(1+\eps)}_{(s,t)})$.
Note that $\dist(s,p_i(s)) = \dist(s,p_i(s), T(p_i(s)))$ and $\dist(t, p_i(s)) = \dist(t, p_i(s), T(p_i(s)))$.
Hence $\dist(s,t) \leq \dist(s,p_i(s)) + \dist(t, p_i(s)) = \dist(s,p_i(s), T(p_i(s))) + \dist(t, p_i(s), T(p_i(s)))$.
In addition, by Lemma \ref{lem:stretch-heavy} we have $\dist(s,t) \leq \hd(s,t) \leq (1+\eps)
\min(\Heap^{(1+\eps)}_{(s,t)})$.

We are left with showing the second direction, namely, $\hat{d}(s,t) \leq (1+\eps)^c(\dist(s,t) +2)$.
Let $P(s,t)$ be the shortest path from $s$ to $t$.
Let $j$ be the index such that $2^{j} \leq \dist(s,t) \leq 2^{j+1}$.
By Lemma \ref{lem:heavy}(2), $S_j$ contains a node $z$ in $P(s,t)$  at distance at most $\eps 2^{j}$ from $s$.

If $T(z)$ does not contain $s$ then by Lemma \ref{lem:not-in-light}, we have
$\Heap^*_Q(s,z) \leq (1+\eps)^2\dist(s,z)$.
We get that $\min(\Heap^{(1+\eps)}_{(s,z)}) \leq \Heap^*_Q(s,z) \leq (1+\eps)^2\dist(s,z)$.
Hence $(1+\eps)\min(\Heap^{(1+\eps)}_{(s,t)}) \leq (1+\eps)^3\dist(s,z)$.

So assume that $T(z)$ contains $s$.
It follows from the definition of the pivot that $\dist(s,p_j(s), T(p_j(s))) \leq \eps 2^{j}$.

Let $v_j = p_j(s)$.
If $T(v_j)$ contains $t$ then we have
$\dist(s,v_j, T(v_j)) + \dist(t,v_j, T(v_j)) = \dist(s,v_j) + \dist(t,v_j) \leq  \eps 2^{j} + \eps 2^{j} + \dist(s,t) = (1+2\eps)\dist(s,t)$.

If $T(v_j)$ does not contain $t$ then by Lemma \ref{lem:not-in-light}, we have
$\Heap^*_Q(t,v_j) \leq (1+\eps)^2\dist(t,v_j)$.
This means that $\dist_Q(t,v_j) \leq \Heap^*_Q(t,v_j) \leq (1+\eps)^2\dist(t,v_j)$.

Let $q \in Q$ be the node that obtains $\Heap^*_Q(t,v_j)$, namely, the node $q$ such that $\Heap^*_q(t,v_j) = \Heap^*_Q(t,v_j)$.

We get that $\min(\Heap^{(1+\eps)}_{(s,t)}) \leq  \Heap^{(1+\eps)}_{(s,t)}[q] \leq (1+\eps)(\dist_q(s,t)) \leq (1+\eps)(\dist(s,v_j) + \dist_q(t,v_j)) \leq
(1+\eps)(\eps 2^j + (1+\eps)^2\dist(t,v_j)) \leq (1+\eps)(\eps 2^j + (1+\eps)^2 (\eps 2^j + \dist(s,t))) \leq
(1+\eps)(\eps 2^j + (1+\eps)^2 (\eps 2^j + \dist(s,t))) \leq (1+\eps)^5 \dist(s,t)$

\QED

\subsection{Reducing the Query Time to $O(1)$}
We now explain how to reduce the query time to $O(1)$.
To get an initial estimation, we use the decremental algorithm of Bernstein and Roditty \cite{BeRo11} with parameter $k=2$
(choosing any constant parameter $k \geq 2$ is sufficient for our needs).
This algorithm has a total update time of $\hat{O}(n^{5/2})$ and can return a distance estimation within a stretch 3.
We can now use the rough estimation  to find the minimal index $i$ such that $t \in T(p_i(s))$.
It is not hard to verify that there are only $O(1)$ potential indices to check.

\appendix

\section{Missing proofs}\label{sec:missing}

\Proofof{Lemma \ref{lem:emulator-very-long-dis}}
$\dist(x,z,H) + \dist(y,z,H) \leq (1+\eps/2)\dist(x,z) + \zeta + (1+\eps/2)\dist(y,z) + \zeta = (1+\eps/2)(\dist(x,z) + \dist(y,z)) + 2\zeta \leq
(1+\eps/2)(2\beta + \dist(x,y)) + 2\zeta \leq (1+\eps/2)\dist(x,y)) + 2\zeta +2\beta(1+\eps/2) =
(1+\eps/2)\dist(x,y) + \eps\beta +2\beta(1+\eps/2) =
(1+\eps/2)\dist(x,y) + 2\beta + 2\eps\beta \leq
(1+\eps/2)\dist(x,y) + 4\beta \leq (1+\eps/2)\dist(x,y) + \eps/2\dist(x,y)  = (1+\eps)\dist(x,y)$.
\QED

\Proofof{Lemma \ref{lem:heavy}}
Consider a fixed graph $G'$.
We show that each event (1)-(3) happens with probability at least $1- 1/n^{c-1}$. The lemma then follows by union bound on all three event.

To see event (1): consider a node $v$, the probability that $\Gamma(v) \cap Q = \emptyset$ is
$\Pr[\Gamma(v) \cap Q = \emptyset] \leq (1- c \ln{n}/n^{1/2})^{n^{1/2}} \leq (1/e)^{c \ln{n}} = 1/n^c$.
By Union Bound on all heavy nodes we get that with probability $1/n^{c-1}$ the Lemma holds.

To see event (2): consider a node $v$ and index $i$ such that there exists a node $z$ such that $\dist(v,z,G') \geq \eps 2^i$.
Note that there are at least $\eps 2^i$ nodes at distance at most $\eps 2^i$ from $v$, namely, $B(v,\eps 2^i,G') \geq \eps 2^i$.

The probability that none of the nodes in $B(v,\eps 2^i,G')$ was selected to $S_i$ is $(1-c\ln{n}/(\eps 2^i))^{\eps 2^i} < n^{-c}$.
By Union Bound on all nodes we get that with probability $1/n^{c-1}$ the Lemma holds.

To see event (3): consider a node $v$  such that $|B(v,\beta)| \geq n^{1/2}$, the probability that $Q \cap B(v,\beta,G') \neq \emptyset$ is
$\Pr[Q \cap B(v,\beta,G') \neq \emptyset] \leq (1- c \ln{n}/n^{1/2})^{n^{1/2}} \leq (1/e)^{c \ln{n}} = 1/n^c$.

By Union Bound on all heavy nodes we get that with probability $1- 3/n^{c-1}$ properties $(1)-(3)$ hold for $G'$.

The random sets $Q$ and $S_i$ are independent of the graph, the failure probability needs to be multiply by the number of considered graphs during the
entire running of the algorithm. Note that as there are $m \leq n^2$ deletions, and thus at most $n^2$ different versions of the graph.
By the union bound on all considered graphs the lemma follows.
\QED

\Proofof{Claim \ref{claim:maintining-exact}}
By Claim \ref{claim:Q-size} the expected size of $Q$ is $\tilde{O}(n^{1/2})$.
For every toe nodes $x,y \in V$ and node $q \in Q$.
The distances $\dist(q,x)$ or $\dist(q,y)$ can be increased at most $8 \beta/\eps$ times before exceeding $8 \beta/\eps$.
Hence $\Heap^{(1)}_{(x,y)}[q]$ is updated at most $O(\beta/\eps) = \hat{O}(1)$ time.
Therefore for all nodes $q \in Q$ updating $\Heap^{(1)}_{(x,y)}$ takes $\hat{O}(n^{1/2})$ total time.
Hence for all pairs $x,y \in V$ updating ${\cal H}^{1}$ takes $\hat{O}(n^{5/2})$ total time.
\QED

\Proofof{Lemma \ref{lem:stretch-heavy}}
Let $P_{heavy}(x,y)$ be the the shortest path from $x$ to $y$ that goes through some heavy node $z$.
Recall that by Lemma \ref{lem:heavy}(1) w.h.p. we have that $Q \cap \Gamma(z) \neq \emptyset$.
Let $z_1\in Q \cap \Gamma(z)$.
Note that $\dist_{z_1}(x,y) \leq \dist_{z}(x,y) + 2 = \hd(x,y) +2 = \dist(x,z) + \dist(z,y) +2$.

Let $z_2 \in Q$ be the node such that $\Heap^{(1+\eps)}_{(x,y)}[z_2] = \min(\Heap^{(1+\eps)}_{(x,y)})$.

We claim that $\dist_{z_2}^{BR}(x,y) \leq (1+\eps) Heap^{(1+\eps)}_{(x,y)}[z_2]$.
To see this, recall that $Heap^{(1+\eps)}_{(x,y)}[z_2]$ is updated when either $\dist^{BR}(z_2,x)$ or $\dist^{BR}(z_2,y)$ increases by a factor of $1+\eps$.

We get that $\dist(x,y) \leq \dist_{z_2}(x,y) \leq \dist_{z_2}^{BR}(x,y) \leq Heap^{(1+\eps)}_{(x,y)}[z_2](1+\eps)$.
In addition, note that $\dist(x,y) \leq \min(\Heap^{(1)}_{(x,y)})$. To see this recall that either $\min(\Heap^{(1)}_{(x,y)}) = \dist_Q(x,y) \geq \dist(x,y)$ or
$\min(\Heap^{(1)}_{(x,y)}) = \infty$. It follows that $\dist(x,y) \leq \tilde{d}(x,y)$.

We left to show the other direction, namely, $\tilde{d}(x,y) \leq (1+\eps)^2(\hd(x,y)+2)$.

If $\dist_Q(x,y) \leq 8 \beta/\eps$ then recall that $\min(\Heap^{(1)}_{(x,y)})) = \dist_Q(x,y)$.
Hence $\tilde{d}(x,y) \leq \min(\Heap^{(1)}_{(x,y)})  = \dist_Q(x,y) = \hd(x,y)+2$.

Otherwise, $\tilde{d}(x,y)\leq  \Heap^{(1+\eps)}_{(x,y)}[z_2](1+\eps) \leq \Heap^{(1+\eps)}_{(x,y)}[z_1](1+\eps) \leq (1+\eps)(1+\eps)\dist_{z_1}(x,y) \leq  (1+\eps)^2 (\hd(x,y)+2)$.
\QED

\Proofof{Lemma \ref{claim:maintain-heavy}}
Consider nodes $x \in V$ and $q \in Q$.
Note that the value $d_{last}(x,q)$ can change at most $\log{n}$ times.
Each time the value $d_{last}(x,q)$ changes, all heap $\Heap^{(1+\eps)}_{(x,y)}$ are updated.
Updating a single heap takes $\tilde{O}(1)$ time.
As there are $n$ such heaps $\Heap^{(1+\eps)}_{(x,y)}$, updating all heaps takes $\tilde(n)$ time.

By Claim \ref{claim:Q-size} the expected size of $Q$ is $\tilde{O}(n^{1/2})$.
We get that the total update time for updating $\Heap^{(1+\eps)}_{(x,y)}$
for all $y\in V$ as a result of a change of $d_{last}(x,q)$ for some $q \in Q$
is $\tilde{O}(n^{3/2})$.

Therefore, the total time for maintaining all heaps $\Heap^{(1+\eps)}_{(x,y)}$ is $\tilde{O}(n^{5/2})$.
\QED

\Proofof{Lemma \ref{claim:maintain-heavy-2}}
Consider pair of nodes $x\in V$, $y \in S_i$ for some $1\leq i \leq \log{n}$.
Maintaining $\Heap^{(*,1+\eps)}_{(x,y)}$ takes $\tilde{O}(2^i n^{1/2})$ total time.
To see this, note that $\Heap^*_q(x,y)$ for some $q\in Q$ is updated every time $\dist^{BR}(q,x)$ or $\dist^{BR}(q,y)$ increases until one of them becomes larger than $2^i$.
This could happen at most $2^i$ times. We get that maintaining $\Heap^{(*,1+\eps)}_{(x,y)}$ takes $\tilde{O}(2^i n^{1/2})$ total time.
There are $\tilde{O}(n/2^i)$ expected number of nodes in $S_i$.
Thus maintaining all heaps $\Heap^{(*,1+\eps)}_{(x,y)}$ for some nodes $x\in V$, $y \in S_i$ takes $\tilde{O}(n \cdot n/2^i \cdot 2^i n^{1/2}) = \tilde{O}(n^{5/2})$.
There are $\log{n}$ indices $i$, therefore maintaining all heap $\Heap^{(*,1+\eps)}_{(x,y)}$ takes $\tilde{O}(n^{5/2})$.
\QED

\Proofof{Lemma \ref{claim:maintain-king}}
By Lemma \ref{lem:heavy}(3) for every node $v$ such that $|B(v,\beta)| \geq n^{1/2}$, $Q \cap B(v,\beta,G) \neq \emptyset$.
In other words, if $\dist(v,Q) > \beta$ then $|B(v,\beta)| < n^{1/2}$.
Note also that the degree of the nodes in $B(v,\beta-1)$ is $O(n^{1/2})$ as $|B(v,\beta)| < n^{1/2}$.
Thus maintaining the distances from $v$ to all nodes in $B(v,\beta)$ using King's algorithm  \cite{King99} takes $\tilde{O}(\beta n^{1/2} \cdot n^{1/2}) = \hat{O}(n^{2})$.
Thus maintaining all $B(v,\beta)$ for all nodes $v$ such that $\dist(v,Q) > \beta$ takes $ \hat{O}(n^{2})$ total time.
\QED

\Proofof{Lemma \ref{lem:short}}
It is not hard to verify that the distances in $T(s)$ are the distances in the induced graph on $V(T(s))$.
Therefore, we need to show that if $v \in T(S)$ then $P(v,s) \subseteq T(s)$.
Or in other words, if there exists a node $x \in P(v,s)$ such that $x \notin T(s)$ then $v$ is not light for $(s,T(s))$.
Let $x$ be the first node on the path $P(v,s)$ such that $x \notin T(s)$.

Namely, $x$ is not light for $(s,T(s))$.
We need to consider the different cases why $x$ is not light for $(s,T(s))$ and show that in each such case $v$ is not light for $(s,T(s))$ as well.

Case $(1)$ is when $\dist(x,s) \leq 8\beta/\eps$ and $\dist(x,s) \geq \dist_Q(x,s) -2$.
Case $(2)$ is when $\dist(x,s) > 8\beta/\eps$, $\dist(x,Q) > 8\beta/\eps$ and $\dist(x,s) \geq \dist(x,w) + \min(\Heap^{(*,1+\eps)}_{(w,s)})/(1+\eps)$ and $w\notin T$ for some $w \in B(x,\beta)$.
Case $(3)$ is when $\dist(x,s) > 8\beta/\eps$ and $\dist(x,s) \geq \min(\Heap^{(*,1+\eps)}_{(x,s)})/(1+\eps)$.

%

Let $H$ be the emulator in the construction of Bernstein and Roditty \cite{BeRo11}.
For a node $q\in Q$, let $H_q$ be the graph $H$ when $Heap^{(1+\eps)}_{(v,s)}[q]$ was last updated.

We now turn to the first case where $\dist(x,s) \leq 8\beta/\eps$ and $\dist(x,s) \geq \dist_Q(x,s) -2$.
We consider two subcases, the first subcase $(1.1)$ is when $\dist(v,s) \leq 8\beta/\eps$ and the second subcase $(1.2)$ is when $\dist(v,s) > 8\beta/\eps$.
In both subcases we have $\dist(v,s) = \dist(v,x) + \dist(x,s) \leq \dist_Q(v,s) + 2$.
Let $q \in Q$ be the node such that $\dist_q(v,s) = \dist_Q(v,s)$.

In case $(1.1)$ since $\dist(v,s) \leq \dist_Q(v,s) + 2$ and $\dist(v,s) \leq 8\beta/\eps$ then $v$ is not light for $(s,T)$ due to check (1).
In case $(1.2)$, we have $\min(\Heap^{(*,1+\eps)}_{(v,s)}) \leq
\dist_q(v,s,H_q) \leq  \dist(v,x,H_q) + \dist(x,q,H_q) + \dist(q,s,H_q) \leq
(1+\eps/2)\dist(v,x) +\zeta + (1+\eps/2)\dist(x,q) +\zeta + (1+\eps/2)\dist(q,s) +\zeta \leq
(1+\eps/2)\dist(v,s) + (1+\eps/2)2 + 3\zeta < (1+\eps)\dist(v,s)$.
So $v$ does not pass check $(3)$.

%
%
%

%
%
%
%
%

Consider case $(2)$ where $\dist(x,s) > 8\beta/\eps$, $\dist(x,Q) \geq \beta$ and $\dist(x,s) \geq \dist(x,w) +  \min(\Heap^{(*,1+\eps)}_{(w,s)})/(1+\eps)$ and $w\notin T$ for some $w \in B(x,\beta)$.

If $v$ has a node $q$ in $Q$ at distance $\beta$ from it then by Lemma \ref{lem:emulator-very-long-dis} we get that
$\min(\Heap^{(*,1+\eps)}_{(v,s)}) \leq \Heap^{(*,1+\eps)}_{(v,s)}[q] = \dist_q(v,s,H_q) \leq (1+\eps)\dist(v,s)$.
We get that $v$ does not pass test (3).

So assume that $\dist(v,Q) > \beta$.
Let $q \in Q$ be the node that $\Heap^{(*,1+\eps)}_{(w,s)}[q] = \min(\Heap^{(*,1+\eps)}_{(w,s)})$. Note that $\dist(v,q) > \beta$.
We have $\dist_q(v,s,H_q) \leq  \dist(v,w,H_q) + \dist_q(w,s,H_q) \leq (1+\eps)(\dist(v,x) + \dist(x,w)) +
\min(\Heap^{(*,1+\eps)}_{(w,s)}) \leq (1+\eps)\dist(v,x) + (1+\eps)\dist(x,s) =
(1+\eps)\dist(v,s)$.
It follows that $v$ does not pass test $(3)$

The last case is when $\dist(x,s) > 8\beta/\eps$ and $\dist(x,s) >  \min(\Heap^{(*,1+\eps)}_{(x,s)})/(1+\eps)$.
If $v$ has a node $q$ in $Q$ at distance $\beta$ from it then similar to the analysis in previous case we can show that $v$ does not pass test $(3)$.
So assume that $\dist(v,Q) > \beta$.

If $\dist(v,x) \leq \beta$ then we get that
$\dist(v,x) +  \min(\Heap^{(*,1+\eps)}_{(x,s)})/(1+\eps)\leq \dist(v,x) + \dist(x,s) = \dist(v,s)$.
Hence, $v$ does not pass test $(2)$.

Assume $\dist(v,x) > \beta$.
In this case, $\min(\Heap^{(*,1+\eps)}_{(v,s)}) = \dist(v,s,H) \leq \dist(v,x,H) + \dist(x,s,H) \leq (1+\eps)\dist(v,x) + (1+\eps)\dist(x,s) = (1+\eps)\dist(v,s)$.
It follows that $v$ does not pass test $(3)$.
\QED

\Proofof{Lemma \ref{lem:nine-ok}}
We claim that
for a node $v \in V$, checking if $v$ is not light for $(s,T(s))$ takes $O(n^{1/2})$ time in expectation.

First note that check (1) and (3) can be done in constant time.
In addition, note that if $v$ is heavy then automatically it is not light for $(s,T(s'))$ for any tree $T(s')$ for some $s' \in S_j$ for $1\leq j \leq \log{n}$.
To see this, recall that by Lemma \ref{lem:heavy} if is heavy then $\Gamma(v) \cap Q \neq \emptyset$.
Let $q \in \Gamma(v) \cap Q $.
If $d(v,s,B(T,1)) \leq 8\beta/\eps$ then note that $\dist_Q(v,s) \leq \dist_q(v,s) \leq \dist(v,s) +2$ and thus $v$ is not light for $(s,T)$ due to test (1).

If $d(v,s,B(T,1)) > 8\beta/\eps$ then straight forward calculations show that $d(v,s,B(T,1)) \geq \dist(v,s) \geq \Heap_Q^*(v,s)/(1+\eps)$.
Thus $v$ is not light for $(s,T)$ due to test (3).
It follows that in case $v$ is heavy it does not pass the property check of $(s,T(s))$.

By Lemma \ref{lem:heavy}(3)
if $\dist(v,Q) \geq \beta$ then $|B(v,\beta)| \geq n^{1/2}$.
Thus checking if
$d(v,s,B(T,1)) \geq \dist(v,w) + \Heap^*(w,s)/(1+\eps)$ for some $w \in B(s,\beta)$ takes $O(n^{1/2})$ time.

We claim that for the tree $T(s)$, the algorithm invokes $O(2^i)$ the check if $v$ is not light for $(s,T(s))$.
To see this, note that the check if $v$ is not light for $(s,T(s))$
is invoked when either the distance $\dist(v,s,T(s))$ increases or when
the distance the distance $\Heap_Q^*(s,v)$ increases. Since we maintain the distances $\dist(v,s,T(s))$ and $\Heap_Q^*(s,v)$ up to depth $2^i (1+\eps)$, we get that these may increase at most $O(2^i)$ times.
In addition, the algorithm go over the edges of $v$ at most $2^i$ times since each time the algorithm go over $v$'s edges then the distance $\dist(v,s)$ increases.
As mentioned before, $T(s)$ contains only light nodes as all heavy nodes are not light for $(s,T(s))$.
Hence maintaining $T(s)$ takes $O(n n^{1/2} 2^{i}) = O(n^{3/2} 2^{i})$.
In expectation there are $\tilde{O}(n/2^i)$ nodes in $S_i$, thus maintaining all trees $T(s)$ for all nodes $s \in S_i$ takes $\tilde{O}(n^{5/2})$ time.
There are $\log{n}$ indices $i$, therefore maintaining all trees $T(s)$ for all nodes in $S_j$ for some $1\leq j \leq \log{n}$ takes $\tilde{O}(n^{5/2})$ total time.
\QED

\end{document}